\documentclass[amssymb,aps,epsf,multicol]{revtex4}
\usepackage{graphicx}

\newcommand{\half}{{\frac{1}{2}}}
\newcommand{\ket}[1]{|#1\rangle}
\newcommand{\bra}[1]{\langle#1|}
\newcommand{\braket}[2]{\langle#1|#2\rangle}

\newcommand{\Buzek}{Bu\v{z}ek }

\newcommand{\mc}{\mathrm}

\begin{document}

\title{Quantum state restoration by quantum cloning and measurement}
\author{K. Maruyama and P. L. Knight}
\affiliation{QOLS, Blackett Laboratory,
Imperial College London, Prince Consort Road, London SW7 2BW, United Kingdom}
\date{\today}

\begin{abstract}
Copying information is an elementary operation in classical
information processing. However, copying seems rather different in
the quantum regime. Since the discovery of the universal quantum
cloning machine, much has been found from the fundamental point of
view about quantum copying. But a basic question as to the utility
of universal quantum cloning remains. We have considered its
application in quantum state restoration by using cloning circuit
for state estimation. It might be expected that classical
information from the state estimation might help restore the
quantum state that was disturbed during storage in a quantum
memory or transmission. We find that the fidelity of the final
state is, interestingly, independent of error probabilities inside
the memory/channel. However, this also turns out to impose a
severe constraint on our original aims.
\end{abstract}
\maketitle

\section{Introduction}
Quantum cloning has been studied intensively \cite{buzek96,
gisin97, gisin98, werner98} and has played an important role in
the development of the theory of quantum information. As copying
information is one of the most fundamental processes in classical
information processing, there has been some hope that quantum
cloning may well be a useful operation in quantum information
processing. However, only a few examples of its practical use have
been discussed (See refs \cite{bechmann99} and \cite{galvao00},
for example).

We attempt first to utilize universal quantum cloning machine
(UQCM) \cite{buzek96, buzek97} in restoring quantum states that
are disturbed during quantum data storage in a quantum memory or
transmission through a noisy quantum channel. Naturally, we have
(approximate) quantum error-correction in mind as a further goal.
By quantum cloning, we wish to reduce the redundancy which is
necessary in both classical and quantum error-correcting schemes,
because using many quantum channels might be expensive in
resources. We have both quantum data storage and quantum
communication in mind. However, we will use a
communication-oriented view with Alice (sender) and Bob
(receiver), which is common in the field of quantum information,
in order to simplify the discussion. If we need to consider the
data storage, we simply interpret Alice's role as the writer of
data and Bob's role as the reader, who restores the quantum state
for the subsequent processing.

\begin{figure}
 \begin{center}
  \includegraphics[scale=0.7]{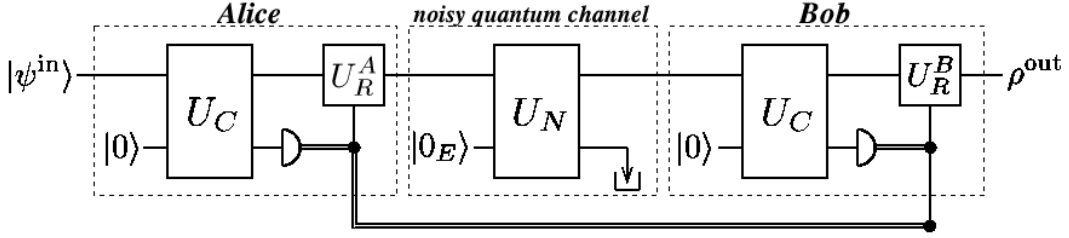}
  \caption{Quantum circuit representation of our state restoration protocol.
Unitary transformation $U_C$ is a universal quantum cloning network with
two extra qubits, which are represented by a single line in this figure.
These extra qubits are observed after being processed with $U_C$ and
the outcome is transmitted through a classical channel, which is
a double line in the figure.
Single qubit gates $U_R^A$ and $U_R^B$ are reversing operations and
Bob's error-correcting operation is included in $U_R^B$. These two
operations are conditional on the outcome of the measurement.
The effect of noise in the transmission channel is represented by
a unitary transformation $U_N$ acting on both the state and its environment.}
 \label{config}
 \end{center}
\end{figure}

Our basic strategy is depicted in Fig. \ref{config} and is as
follows. By measuring two out of three qubits emerging from a
quantum cloning circuit, Alice obtains some information about the
initial state and sends this information to Bob using a classical
channel. After transmitting the state through a noisy quantum
channel, Bob also acquires information on the received state in
the same fashion. If there is no energy dissipation during the
transmission, Bob may be able to infer what kind of error has
affected the state by comparing his measurement results with those
of Alice. Then, he can apply the inverse of error operations to
make the state as close as possible to the initial state.

One advantage of this idea is that it may work even if the error rates
are very high. This is because Bob infers the type of error by measurement
results, instead of through the error syndrome which relies on redundancy,
so there is no need to assume low error probabilities, which are common in
the standard error-correcting methods.

As well as utilizing the quantum cloning transformation, we also introduce
an operation to reverse the effect of quantum measurement in order to
improve the fidelity. Quantum measurements are, of course, irreversible,
thus this reversal can be performed only approximately, when we wish this
reversal to be a deterministic process.

We have found that the fidelity between initial and final states does not
depend on the error probabilities during transmission. This can be seen as a
consequence of the universality of the quantum cloning transformation. However,
this feature imposes some constraint on the fidelity and our scheme exemplifies
a situation in which use of both classical and quantum channels are not
necessarily sufficient to improve the final fidelity.

This paper is organized as follows. In Section \ref{sec_st_es}, we describe how a
quantum cloning circuit can be used to estimate quantum states.
Then, in Section \ref{sec_protocol}, we explain the overall protocol.
Section \ref{sec_rev_op} shows the approximate reversal of quantum
measurement. Results and some discussions are given in Section \ref{sec_results}
and concluding remarks in Section \ref{sec_remarks}.


\section{State estimation by cloning}\label{sec_st_es}
As in most literature on quantum error-correction
\cite{steaneknill}, here we take only bit and phase flips into
account as the types of errors that may occur in the quantum
channel. Thus, we wish to detect these two types of errors, whose
occurrence may be described with two bit information, by comparing
Alice's and Bob's measurement results. Since measurement on two
qubits gives two bit classical information at most, it may be
enough if the measurement can give the information on the tendency
about the bit and the phase, i.e. one bit information for each. As
we do not assume any a priori knowledge about the state, what we
need to estimate is a quadrant of the space where the sate
resides. Estimating the most probable quadrant for an incoming
state by a UQCM circuit proceeds as follows. Let us consider
Alice's case, as Bob's estimation is performed exactly the same
way.

The initial state $\ket{\psi^{\mc{in}}}$ can be written as
\begin{equation} \label{instate}
 \ket{\psi^{\mc{in}}}=\alpha\ket{0}+\beta e^{i\phi}\ket{1},
\end{equation}
where $\alpha, \beta \ge0 \: (\alpha^2+\beta^2=1)$ by neglecting
the unimportant global phase. We consider only pure states as
input for simplicity. The output state from Alice's cloning
transformation is
\begin{equation} \label{Psi_Aout}
 \ket{\Psi_A^{\mc{cloned}}}=\sqrt{\frac{2}{3}}(\alpha\ket{000}
+\beta e^{i\phi}\ket{111}) + \sqrt{\frac{1}{6}}\left(\alpha(\ket{011}+\ket{101})
+\beta e^{i\phi}(\ket{010}+\ket{100})\right).
\end{equation}

\begin{figure}
 \begin{center}
  \includegraphics[scale=0.35]{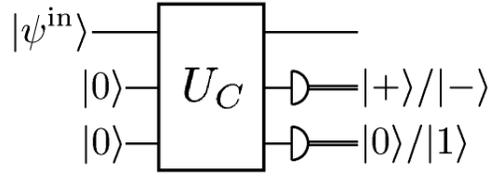}
  \caption{Measurements on qubits from a cloning circuit. The second and the
third qubits are measured using basis sets $\{\ket{+},\ket{0}\}$ and
$\{\ket{0},\ket{1}\}$, respectively.}
 \label{Ucs}
 \end{center}
\end{figure}

\begin{figure}
 \begin{center}
  \includegraphics[scale=0.6]{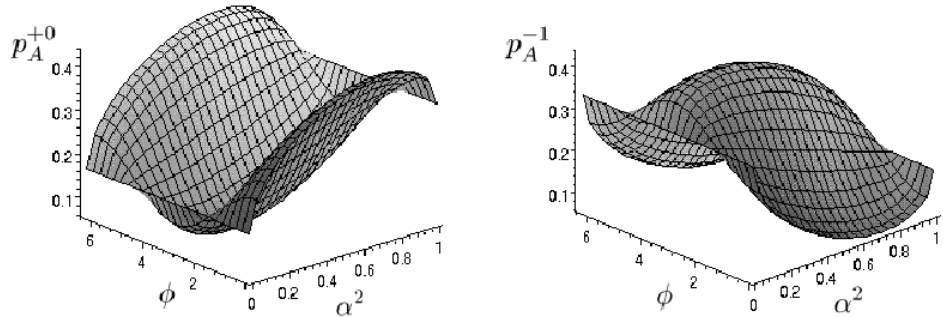}
  \caption{The probability distributions $p_A^{+0}$ and $p_A^{-1}$.
It is easily seen that $p_A^{+0}$ is relatively higher where $\cos\phi>0$
and $\alpha^2>\half$ and $p_A^{-1}$ has the opposite tendency.
The other probability distribution, such as $p_A^{+1}$, is merely
a combination of $\phi$-dependence of $p_A^{+0}$ and $\alpha^2$-dependence
of $p_A^{-1}$.}
 \label{p_A}
 \end{center}
\end{figure}

In order to obtain information on $\alpha, \beta$ and $\phi$, Alice
performs projective measurements on the second and the third qubits
in terms of basis sets \{$\ket{\pm}=\frac{1}{\sqrt{2}}(\ket{0}\pm\ket{1})$\}
and \{$\ket{0}, \ket{1}$\}, respectively, as shown in Fig. \ref{Ucs}.
The probability of having the outcome of $\ket{+}$ from the measurement
on the second qubit of the state above is
\begin{equation}\label{paplus}
 p_{A(2)}^+=\half\left(1+\frac{4}{3}\alpha\beta\cos\phi\right),
\end{equation}
where the subscript $A(2)$ stands for Alice's second qubit, and the
probability of having $\ket{-}$ is, thus, $p_{A(2)}^-=\half\left(1-
\frac{4}{3}\alpha\beta\cos\phi\right)$. Since we are assuming that
both $\alpha$ and $\beta$ are non-negative, if $p_{A(2)}^+ > p_{A(2)}^-$
then $\cos\phi >0$ and if $p_{A(2)}^+ < p_{A(2)}^-$ then $\cos\phi <0$. Therefore,
by interpreting the outcome $\ket{+}$ (or $\ket{-}$) as a consequence
of the relation between probabilities, $p_{A(2)}^+ > p_{A(2)}^-$
(or $p_{A(2)}^+ < p_{A(2)}^-$), Alice can estimate the range
$\phi$ is in with a high probability.

Similarly, the probabilities of obtaining $\ket{0}$ and $\ket{1}$ from
the measurement on the third qubit are
$p_{A(3)}^0=\frac{1}{3}(1+\alpha^2)$ and $p_{A(3)}^1=\frac{1}{3}(1+\beta^2)$,
respectively; hence we can say that the outcome $\ket{0}$ implies
$\alpha>\beta$ and $\ket{1}$ implies the opposite.
Fig. \ref{p_A} shows a joint probability $p_A^{+0}$, which is the probability
of outcomes + for the second qubit and 0 for the third qubit, and also
$p_A^{-1}$ for comparison. Each probability distribution's dependence on
$\alpha^2$ and $\phi$ can be easily seen in this figure.

It is also worth noting for our protocol that the measurement described
above does not change the tendency concerning $\alpha, \beta,$ and
$\cos\phi$, as long as the implications are correct. For example, the
state after Alice obtains + and 0 from her measurement can be written as
\begin{eqnarray}
\ket{\psi_A^{+0}} &=& \frac{1}{||\cdot||}\left((\alpha+\half\beta e^{i\phi})\ket{0}
+ \half\beta e^{i\phi}\ket{1}\right) \nonumber \\
&=& \frac{1}{||\cdot||}(\alpha'\ket{0}+\beta'e^{i\phi'}\ket{1}),
\end{eqnarray}
where
\begin{eqnarray}
\alpha' = \left(\alpha^2+\alpha\beta\cos\phi+\frac{1}{4}\beta^2\right)^\half,
\beta' = \half\beta, \nonumber \\
\phi' = \phi-\tan^{-1}\frac{\beta\sin\phi}{2\alpha+\beta\cos\phi},
\end{eqnarray}
and the global phase is included in the normalization factor, which is denoted
by $||\cdot ||$ symbol. If the implications by the measurement are correct,
i.e., $\alpha>\beta$ and $\cos\phi>0$, then $\alpha', \beta'$, and $\phi'$
still satisfy the same relationship, $\alpha'>\beta'$ and $\cos\phi'>0$.

\section{The protocol}\label{sec_protocol}
Our task here is to send an unknown pure state
$\ket{\psi^{\mc{in}}}$, a \textit{signal state}, through a noisy
channel with as high fidelity as possible. We take the ordinary
input-output fidelity, which is defined by
$F=\bra{\psi^{\mc{in}}}\rho^{\mc{out}}\ket{\psi^{\mc{in}}}$ with a
density matrix of the output state $\rho^{\mc{out}}$, as a figure
of merit of the protocol's performance. This means that our
protocol is rather different from other quantum error-correcting
schemes where we need to maintain the entanglement fidelity close
to unity \cite{ent_fid}.

Let us now describe the protocol in detail. In Fig. \ref{config},
$U_C$ is a quantum circuit implementing $1\rightarrow 2$ universal
quantum cloning transformation. Since a $1\rightarrow 2$ UQCM
requires one ancilla qubit to produce two output qubits, $U_C$
processes three qubits and outputs an entangled three-qubit state.
In the figure, the second and third qubits are represented by a
single line.

Alice lets the initial state $\ket{\psi^{\mc{in}}}$ go through a
UQCM circuit and performs measurements on the second and third qubits emerging
from the circuit to obtain some information on $\ket{\psi^{\mc{in}}}$.
As these measurements, of course, disturb the signal state, Alice
tries to make it as close to the initial state as possible. The reversal
operation, $U_R^A$, should be performed deterministically depending on the
outcome of the measurement, so it is a conditional unitary operation upon
the outcome as a control bit. This is represented by a unitary gate connected
with the classical information channel in Fig. \ref{config}. A filled
black circle denotes a control bit.
We will describe the details of the reversal operation later in Section
\ref{sec_rev_op}. Alice also sends the results of her measurements to Bob
through a classical communication channel, which we assume is error-free.

The effect of noise on the signal state during a transmission can be
understood as a result of a unitary transformation $U_N$ acting on the
state and its environment, which can be taken as a pure state $\ket{0_E}$,
and the state of the environment after the transformation
is discarded without being observed. This view leads to the standard
\textit{Kraus representation} of a quantum operation \cite{kraus},
$\mathcal{E}(\rho)=\sum E_i \rho E_i^\dagger$. $E_i$ are operators
acting on the state space of the principal system $\rho$ and they can be in
general written as $E_i=\bra{e_i}U\ket{0_E}$, where $\ket{e_i}$ are the
orthonormal basis for the state space of the environment. If we denote
the probability of a bit flip, which swaps $\ket{0}$ and $\ket{1}$,
by $p_{bit}$ and that of a phase flip, which flips the sign of $\ket{1}$
while that of $\ket{0}$ is unchanged, is $p_{ph}$, the Kraus form of the
error operation becomes
\begin{equation}\label{errorkraus}
 \mathcal{E}^{er}(\rho) = \sum_{i=0}^{4}E_i^{er}\rho {E_i^{er}}^\dagger,
\end{equation}
where $E_0^{er}=\sqrt{(1-p_{bit})(1-p_{ph})}I, E_1^{er}=
\sqrt{p_{bit}(1-p_{ph})}\sigma_x, E_2^{er}=\sqrt{p_{ph}(1-p_{bit})}\sigma_z,$
and $E_3^{er}=\sqrt{p_{bit}p_{ph}}\sigma_x\sigma_z$.

Bob performs the same cloning transformation on the signal state
he receives and measures the second and third qubits to acquire
information on the state. Then, by comparing the outcome of his
measurement with that from Alice, he infers what type of error has
occurred during the transmission and carries out both the reversal
and the error-correcting operations accordingly to output the
final state $\rho^{\mc{out}}$. More specifically, if Alice's and
Bob's outcomes from the measurement on the second qubit disagree,
then Bob infers that there has been a phase flip and applies
$\sigma_x$ to the state. Similarly, if their outcomes for the
third qubit disagree, he flips the bit with $\sigma_z$. If both
disagree, $\sigma_x\sigma_z$ will be applied.

\section{The reversal of quantum measurement}\label{sec_rev_op}
Since a quantum state will be disturbed by any form of quantum measurement
in principle, the signal state after the measurement for state estimation
is no longer the same as the original state. However, in order to make
the final fidelity as close as possible to unity, let us undertake an
attempt to reverse the quantum measurement. Such a reversal, of course,
can never be achieved perfectly, but probabilistic perfect reversals
are possible and have been discussed in the context of quantum error-correction
(\cite{ref_rev_op}, for example). Nevertheless, what we discuss here is
a (deterministic) approximate reversal, because the form of measurement,
including the cloning transformation, is fixed in our case, thus,
there is little freedom to apply the perfect reversal.

It is easily seen from Figs. \ref{config} and \ref{Ucs} that measuring
second and third qubits is equivalent to measuring the state of
the environment, which is provided as a pure state $\ket{0}$ initially,
after a unitary evolution $U_C$. Therefore, each of four measurement
outcomes corresponds to an operation element in the Kraus representation
of the quantum cloning transformation,
$\mathcal{E}_C(\ket{\psi}\bra{\psi})=\sum_i E_i \ket{\psi}\bra{\psi}{E_i}^\dagger$.
Our analysis is similar in approach to the conditional dynamics utilized
in quantum jump analyses of quantum trajectories \cite{martin98}.
Operation elements can be expressed as
\begin{eqnarray}\label{op_elements}
E_0=\frac{1}{2\sqrt{3}}\left( \begin{array}{cc} 2 & 1 \\ 0 & 1 \end{array}
\right), & &
E_1=\frac{1}{2\sqrt{3}}\left( \begin{array}{cc} 1 & 0 \\ 1 & 2 \end{array}
\right), \nonumber \\
E_2=\frac{1}{2\sqrt{3}}\left( \begin{array}{cc} 2 & -1 \\ 0 & 1 \end{array}
\right), & &\mbox{and }
E_3=\frac{1}{2\sqrt{3}}\left( \begin{array}{cc} -1 & 0 \\ 1 & -2 \end{array}
\right),
\end{eqnarray}
where subscripts $\{0,1,2,3\}$ denote measurement outcomes $\{+0, +1, -0, -1\}$.
Hence, if we obtained a measurement outcome $i\in\{0,1,2,3\}$, the signal
state after the measurement becomes
\begin{equation}\label{postmeas}
\ket{\psi'}_i = \frac{E_i\ket{\psi}}
{\sqrt{\bra{\psi}E_i^\dagger E_i\ket{\psi}}}.
\end{equation}
The approximate reversal of quantum measurement can be performed by the inverse
of a unitary operator that is ``similar'' to the non-Hermitian operator, $E_i$.
Thus, the question is simply to find a unitary matrix which is closest to a given
matrix and is independent of the state $\ket{\psi}$, as we
have no preferred input state.
We can choose any metric for matrices to measure similarity between matrices.
Here, we take a metric defined with the Hilbert-Schmidt norm, i.e., we define
the distance between two matrices $A$ and $B$ as
$\mathrm{dist}(A,B)=\left(\mc{Tr}[(A-B)^\dagger(A-B)]\right)^{1/2}$.

Suppose that we are approximating a square matrix $E$. By singular value
decomposition \cite{bhatia}, $E$ can be written as $E=VDW$, where $V$ and $W$
are certain unitary matrices and $D$ is a diagonal matrix with non-negative entries.
Therefore, approximating $E$ by a unitary operation is now
equivalent to approximating $D$ by a unitary. Such a unitary turns out to be
the identity matrix, $I$, as follows. Let us denote a $2\times 2$ diagonal matrix
$D$ by $\left(\begin{array}{cc}\mu & 0 \\ 0 & \nu \end{array}\right)$.
Without loss of generality, we can assume $\mu>\nu$, otherwise we can
simply multiply $\sigma_x$ from both sides and include it in $V$ and $W$.
In general, a $2\times 2$ unitary matrix can be expressed as
$T=\left(\begin{array}{cc}a & b \\ -b^* & a^* \end{array}\right)$,
with $|a|^2+|b|^2=1$, thus, the distance between these two
matrices is given by
\begin{eqnarray}\label{dist}
\mc{dist}(W, D)^2
 &=& \mc{Tr}[(T-\lambda D)^\dagger(T-\lambda D)] \nonumber \\
 &=& 2-\lambda((\mu+\nu)a+(\mu-\nu)a^*)+\lambda^2\nu^2,
\end{eqnarray}
with a certain positive normalization factor $\lambda$, corresponding to the
denominator in Eq. (\ref{postmeas}). To minimize the value of
Eq. (\ref{dist}), $a$ should be equal to 1 as both $\mu$ and $\nu$ are
non-negative. It follows that the closest unitary matrix to a diagonal
matrix with non-negative entries is the identity matrix and the closest
unitary matrix to $E$ is $VW$. Since the singular value decomposition can
be seen as a consequence of the polar decomposition,
$VW$ is equal to the unitary matrix, $U$, that appears in the polar
decomposition of $E$, $E=U\sqrt{E^\dagger E}$ \cite{bhatia2}. Reversal matrices
for the operation elements in Eq. (\ref{op_elements}) are then written as
\begin{eqnarray}\label{rev_elements}
{U_R^0}^\dagger=\frac{1}{\sqrt{10}}\left( \begin{array}{cc} 3 & -1 \\ 1 & 3 \end{array}
\right), & &
{U_R^1}^\dagger=\frac{1}{\sqrt{10}}\left( \begin{array}{cc} -3 & -1 \\ 1 & -3 \end{array}
\right), \nonumber \\
{U_R^2}^\dagger=\frac{1}{\sqrt{10}}\left( \begin{array}{cc} 3 & 1 \\ -1 & 3 \end{array}
\right), & &\mbox{and }
{U_R^3}^\dagger=\frac{1}{\sqrt{10}}\left( \begin{array}{cc} -3 & 1 \\ -1 & -3 \end{array}
\right).
\end{eqnarray}

\begin{figure}
 \begin{center}
  \includegraphics[scale=0.5]{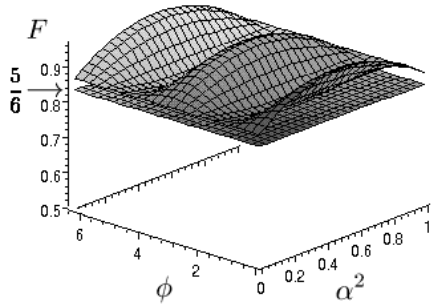}
  \caption{Fidelity after applying the cloning and the reversal operations.
Lower plane shows the ``normal'' fidelity of a $1\rightarrow 2$ UQCM, which is $5/6$.}
 \label{F_rev}
 \end{center}
\end{figure}

Fig. \ref{F_rev} shows the fidelity of the output state after applying the
cloning transformation and the reversal operation. The plane at $F=\frac{5}{6}$
represents the normal fidelity of the output state from a $1\rightarrow 2$
UQCM. The effect of the reversal operation can be seen clearly in the figure:
The fidelity is raised from $\frac{5}{6}$ by using the information form the
measurement.

The reversal operation may change the quadrant in which the signal state lies
in the $\alpha-\phi$ plane. In fact, it does change in some cases and
thus measurement outcomes from Alice and Bob sometimes disagree even if
there was no error and the estimation was perfect.
However, as such a case is rather rare and the fidelity
after the reversal is never lower than the case where we do not apply
the reversal operation as in Fig. \ref{F_rev}, we assume that the reversal
operation does not affect the tendency about $\alpha, \beta$, and $\cos\phi$
so that the signal state stays in the same quadrant as long as the implications
by the measurement are correct.

\section{Results}\label{sec_results}
\begin{figure}
 \begin{center}
  \includegraphics[scale=0.5]{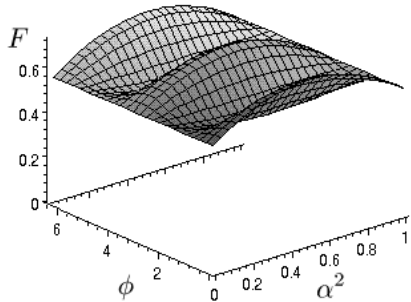}
  \caption{The fidelity between the initial and the final states. The overall
average is 0.593. The value of $F$ is always over $1/2$, except for two points,
$(\alpha^2,\phi)=(1/2,\pi/2)$ and $(1/2,3\pi/2)$.}
 \label{Ffinal}
 \end{center}
\end{figure}

The numerically calculated fidelity between the initial and the final
states is plotted in Fig. \ref{Ffinal}. The average fidelity over the
$\alpha-\phi$ plane is 0.593, which is rather low if we regard the protocol as an
error-correcting scheme. This value is even lower than that of a much
simpler protocol, i.e., a direct measurement by Alice with a basis
$\{\ket{0}, \ket{1}\}$ and the generation of either $\ket{0}$ or
$\ket{1}$ by Bob according to Alice's measurement result.
This protocol gives an average fidelity of $\frac{2}{3}$ and it is
optimal as state estimation \cite{massar95}. If our protocol is
equivalent to state estimation, it is natural to have a fidelity lower
than $\frac{2}{3}$. However, this is not the case, because estimating
the state is different from the purpose of our protocol. All we wish to have
is a high fidelity between the unknown initial state and the unknown final
state. The fidelity can be higher than $\frac{2}{3}$ when the error rates
of the channel is low enough and the measurements by Alice and Bob are weak enough.

\begin{figure}
 \begin{center}
  \includegraphics[scale=0.7]{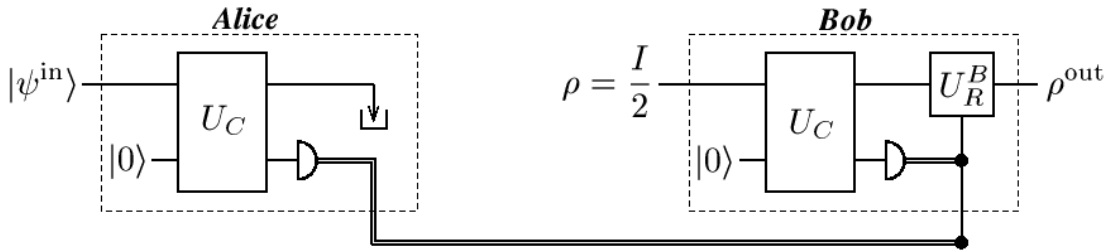}
  \caption{Quantum circuit equivalent to that in Fig. \ref{config} when
the quantum channel is completely noisy. As the state Bob receives is
completely random, it contains no information about the initial
state $\ket{\psi^\mc{in}}$. Thus, the quantum channel is no longer
necessary in such a case.
The signal state is discarded after being processed by Alice's $U_C$ and Bob
follows the same procedure of the protocol after receiving a maximally
mixed state, whose density matrix is given by $I/2$.}
 \label{config_mixed}
 \end{center}
\end{figure}

Nevertheless, the fidelity by our protocol is well below
$\frac{2}{3}$. This is partly because of another interesting
feature: The fidelity does not depend on the error rates,
$p_{bit}$ and $p_{ph}$. That is, a perfect channel
($p_{bit}=p_{ph}=0$) and a completely noisy channel
($p_{bit}=p_{ph}=1/2$) give the same fidelity as in Fig.
\ref{Ffinal}. This means that we can achieve the same value by
providing a maximally mixed state to Bob without using the quantum
channel. A quantum circuit equivalent to such an extreme situation
is depicted in Fig. \ref{config_mixed}. Instead of a randomly
disturbed signal state, a maximally mixed state, whose density
matrix is given by $I/2$ with $I$ representing a $2\times 2$
identity matrix, is generated from a certain source and provided
to Bob as a signal state. Bob performs the same procedure
according to the protocol for the incoming completely mixed state.
In this extreme case, components of our protocol are the same as
those in the simple one whose fidelity reaches $\frac{2}{3}$,
i.e., a measurement on an unknown quantum state, the transmission
of the outcome through a classical channel, and the reproduction
of state using the classical information. Therefore, the fidelity
should be lower than $\frac{2}{3}$. The insensitivity of the
fidelity to error rates means that it is always lower than
$\frac{2}{3}$, regardless of the error rates.

We find it interesting that the completely noisy channel scenario
(Fig. \ref{config_mixed}) gives the same fidelity as the case in which
the quantum channel is completely noise-free, i.e., $U_N=I$ in Fig. \ref{config}.
The information retained in the signal state does not have any effect on
the fidelity in this protocol, as if all information that are \textit{immune}
to errors were absorbed by Alice's measurement through the cloning transformation.

The reason for the fidelity's insensitivity to error rates is in
the symmetry among outputs of $U_C$, stemming from the
universality of the transformation. In order to illustrate this,
let $p_\mc{ne}^{01}$, for example, denote the probability that Bob
obtains $1=\{+1\}$ by his measurement after Alice obtains
$0=\{+0\}$ and there has been no error in the channel; the
superscripts stand for Alice's and Bob's measurement results and
the subscripts indicate the type of error occurred, ne=(no error),
bf=(bit flip), pf=(phase flip) and bpf=(bit and phase flips).
Similarly, let state vectors, such as $\ket{\psi_{ne}^{01}}$,
denote final states in corresponding situations.

In our protocol, for example, $p_\mc{ne}^{00}, p_\mc{bf}^{01}, p_\mc{pf}^{02}$
and $p_\mc{bpf}^{03}$ are equal. Also, a set of final states,
$\ket{\psi_\mc{ne}^{00}}, \ket{\psi_\mc{bf}^{01}}, \ket{\psi_\mc{pf}^{02}}$
and $\ket{\psi_\mc{bpf}^{03}}$ are the same as well. These can be
verified straightforwardly by calculating each probability and state as
\begin{eqnarray}\label{ps_and_psis}
p_\mc{ne}^{00} &=& \bra{\psi_A^0}E_0^\dagger E_0\ket{\psi_A^0}
= \frac{1}{p_A^0}\bra{\psi^\mc{in}}(E_0^\dagger E_0)^2\ket{\psi^\mc{in}},
 \nonumber \\
p_\mc{bf}^{01} &=& \frac{1}{p_A^0}\bra{\psi^\mc{in}}\sqrt{E_0^\dagger E_0}
 \sigma_x E_1^\dagger E_1 \sigma_x\sqrt{E_0^\dagger E_0}\ket{\psi^\mc{in}}
 \nonumber \\
&=& \frac{1}{p_A^0}\bra{\psi^\mc{in}}(E_0^\dagger E_0)^2\ket{\psi^\mc{in}}
= p_\mc{ne}^{00}, \nonumber \\
p_\mc{pf}^{02} &=& \cdots  = p_\mc{ne}^{00},\nonumber \\
p_\mc{bpf}^{03} &=& \cdots  = p_\mc{ne}^{00},\nonumber \\
\ket{\psi_\mc{ne}^{00}} &=&
 \frac{E_0^\dagger E_0\ket{\psi^\mc{in}}}
 {\sqrt{\bra{\psi^\mc{in}}(E_0^\dagger E_0)^2\ket{\psi^\mc{in}}}}
 = \frac{E_0^\dagger E_0 \ket{\psi^\mc{in}}}{\sqrt{p_A^0 p_\mc{ne}^{00}}},
 \nonumber \\
\ket{\psi_\mc{bf}^{01}} &=& \frac{1}{\sqrt{p_A^0 p_\mc{bf}^{01}}}
 \sigma_x\sqrt{E_1^\dagger E_1}\sigma_x\sqrt{E_0^\dagger E_0}\ket{\psi^\mc{in}},
 \nonumber \\
&=& \frac{E_0^\dagger E_0 \ket{\psi^\mc{in}}}{\sqrt{p_A^0 p_\mc{ne}^{00}}}
= \ket{\psi_\mc{ne}^{00}}, \nonumber \\
\ket{\psi_\mc{pf}^{02}} &=& \cdots =\ket{\psi_\mc{ne}^{00}}, \nonumber \\
\ket{\psi_\mc{bpf}^{03}} &=& \cdots =\ket{\psi_\mc{ne}^{00}},
\end{eqnarray}
using specific forms of operation elements in Eq. (\ref{op_elements}).
In Eqs. (\ref{ps_and_psis}), $p_A^0$ denotes the probability for Alice
to have the outcome $0=\{+0\}$ and
$\ket{\psi_A^0}=\sqrt{E_0^\dagger E_0}\ket{\psi^\mc{in}}/\sqrt{p_A^0}$
is the state after Alice's reversal operation.

As a result of the structure of the states and probabilities due to the
symmetry in the quantum cloning transformation, many terms in the fidelity
cancel out each other and all $p_{bit}$ and $p_{ph}$ disappear.
For example, the fidelity after Alice measures $0=\{+0\}$ can be computed as
$F^0=(1-p_{bit})(1-p_{ph})(p_\mc{ne}^{00}|\braket{\psi^\mc{in}}{\psi_\mc{ne}^{00}}|^2
+ \cdots) + p_{bit}(1-p_{ph})(p_\mc{bf}^{01}|
  \braket{\psi^\mc{in}}{\psi_\mc{bf}^{01}}|^2
+ \cdots) + p_{ph}(1-p_{bit})(p_\mc{pf}^{02}|
  \braket{\psi^\mc{in}}{\psi_\mc{pf}^{02}}|^2
+ \cdots) + p_{bit}p_{ph}(p_\mc{bpf}^{03}|
  \braket{\psi^\mc{in}}{\psi_\mc{bpf}^{03}}|^2
+ \cdots) = p_\mc{ne}^{00}|\braket{\psi^\mc{in}}{\psi_\mc{ne}^{00}}|^2
+ \cdots$.

It is not hard to calculate the final fidelity analytically thanks to its
independence on error rates. Assuming $I/2$ as the input state to Bob's circuit,
we obtain
\begin{equation}\label{fidelity_exact}
 F=\frac{1}{9}(5-2\alpha^2+2\alpha^4)+\frac{8}{9}\alpha^2\beta^2\cos^2\phi,
\end{equation}
which reproduces the same plot as Fig. (\ref{Ffinal}). The average
turns out to be $\frac{16}{27}=0.5926$, in accordance with our
numerical result.

\section{Concluding remarks}\label{sec_remarks}
We investigated the possibility of the practical application of
universal quantum cloning in quantum state restoration. We
expected that the classical information from the state estimation
by cloning would help improve the fidelity of the state after
quantum data storage or transmission in a noisy environment. We
have found that the fidelity does not depend on the error
probabilities during transmission, thanks to the universality of
the cloning transformation. However, this feature leads to a lower
fidelity than its optimal value for state estimation of a single
qubit, even if the initial state stays undisturbed when quantum
memory/channel has low error rates. It implies that the
acquisition of both classical and quantum information does not
necessarily improve the fidelity even if nothing is discarded
unnecessarily except for some information loss due to the
interaction with the environment.

Although we have focused on the use of universal quantum cloning
machine and individual measurement on two output qubits from it,
there is a possibility of optimization of the cloning
transformation and a joint measurement. Especially, if the number
of possible input states $\ket{\psi^{\mc{in}}}$ is limited, we may
be able to make use of the state-dependent quantum cloning
\cite{bruss98}. A higher fidelity can be expected in such a case
and it might be ``useful" as an approximate error-correcting
scheme in terms of real cost for implementation. We will discuss
it elsewhere in the future.

\section*{Acknowledgments}
We are grateful to A. Beige for a lot of discussions. This work
was supported in part by the UK Engineering and Physical Sciences
Research Council, the European Union and Fuji Xerox.


\begin{thebibliography}{99}
\bibitem{buzek96}
V. \Buzek and M. Hillery, Phys. Rev. A \textbf{54}, 1844 (1996).
%
\bibitem{gisin97}
N. Gisin and S. Massar, Phys. Rev. Lett. \textbf{79}, 2153 (1997).
%
\bibitem{gisin98}
N. Gisin, Phys. Lett. A \textbf{242}, 1 (1998).
%
\bibitem{werner98}
R. F. Werner, Phys. Rev. A \textbf{58}, 1827 (1998).
%
\bibitem{bechmann99}
H. Bechmann-Pasquinucci and N. Gisin, Phys. Rev. A \textbf{59}, 4238 (1999).
%
\bibitem{galvao00}
E. F. Galv{\~a}o and L. Hardy, Phys. Rev. A \textbf{62}, 022301 (2000).
%
\bibitem{buzek97}
V. \Buzek, S. L. Braunstein, M. Hillery, and D. Bruss, Phys. Rev.
A \textbf{56}, 3446 (1997).
%
\bibitem{steaneknill}
For example, A. Steane, Proc. Roc. Soc. Lond. A \textbf{452}, 2551
(1996); E. Knill and R. Laflamme, Phys. Rev. A \textbf{55}, 900
(1997).
%
\bibitem{ent_fid}
B. Schumacher, Phys. Rev. A \textbf{54}, 2614 (1996); M. A. Nielsen,
\eprint{quant-ph/9606012}.
%
\bibitem{kraus}
K. Kraus, \textit{States, Effects, and Operations} (Springer-Verlag,
Berlin, 1983).
%
\bibitem{ref_rev_op}
M. A. Nielsen, C. M. Caves, B. Schumacher, and H. Barnum, Proc. R. Soc.
Lond. A \textbf{454}, 277 (1998); M. Koashi and M. Ueda, Phys. Rev. Lett.
\textbf{82}, 2598 (1999).
%
\bibitem{martin98}
M. B. Plenio and P. L. Knight, Rev. Mod. Phys. \textbf{70}, 101 (1998).
%
\bibitem{bhatia}
R. Bhatia,
\textit{Matrix Analysis} (Springer-Verlag, New York, 1997).
%
\bibitem{bhatia2}
This is a simple proof of the problem for our purpose. A more formal
proof of the same proposition can be found in \cite{bhatia}.
%
\bibitem{massar95}
S. Massar and S. Popescu, Phys. Rev. Lett. \textbf{74}, 1259 (1995).
%
\bibitem{bruss98}
D. Bru\ss, D. P. DiVincenzo, A. Ekert, C. A. Fuchs, C.
Macchiavello, and J. A. Smolin, Phys. Rev. A \textbf{57}, 2368
(1998).
%
\end{thebibliography}
\end{document}